\documentclass[letterpaper]{article}
\usepackage{setspace}
\usepackage[ascii]{inputenc}
\usepackage[T1]{fontenc}
\usepackage[english]{babel}
\usepackage{amsmath}
\usepackage{amssymb,amsfonts,textcomp}
\usepackage{color}
\usepackage{array}
\usepackage{supertabular}
\usepackage{hhline}
\usepackage{hyperref}
\hypersetup{pdftex, colorlinks=true, linkcolor=blue, citecolor=blue, filecolor=blue, urlcolor=blue, pdftitle=, pdfauthor=coxlab, pdfsubject=, pdfkeywords=}
\usepackage[pdftex]{graphicx}
\newcommand\textsubscript[1]{\ensuremath{{}_{\text{#1}}}}
\makeatletter
\newcommand\arraybslash{\let\\\@arraycr}
\makeatother
\makeatletter
\newcommand\ps@Standard{
  \renewcommand\@oddhead{}
  \renewcommand\@evenhead{}
  \renewcommand\@oddfoot{}
  \renewcommand\@evenfoot{}
  \renewcommand\thepage{\arabic{page}}
}
\makeatother
\title{}
\author{coxlab}
\date{2013-03-29}

\begin{document}
\clearpage\setcounter{page}{1}\pagestyle{Standard}

\textbf{The Convergence of eQTL Mapping, Heritability Estimation and
Polygenic Modeling: \ Emerging Spectrum of Risk Variation in Bipolar
Disorder}

\bigskip

Eric R. Gamazon\textsuperscript{1}, Hae Kyung Im\textsuperscript{2},
Chunyu Liu\textsuperscript{3}, Members of the Bipolar Disorder Genome
Study (BiGS) Consortium\textsuperscript{6}, Dan L.
Nicolae\textsuperscript{1,4}, Nancy J. Cox\textsuperscript{1,5*}

\textsuperscript{1}Section of Genetic Medicine, Department of Medicine,
The University of Chicago

\textsuperscript{2}Department of Health Studies, The University of
Chicago

\textsuperscript{3}Department of Psychiatry, The University of Illinois,
Chicago

\textsuperscript{4}Department of Statistics, The University of Chicago

\textsuperscript{5}Department of Human Genetics, The University of
Chicago

\textsuperscript{6}Members of the BiGS Consortium are listed at the end
of the article.

\bigskip

\bigskip

\bigskip

\bigskip

\bigskip

\bigskip

\bigskip

\bigskip

\bigskip

\bigskip

\bigskip

\bigskip

\bigskip

\bigskip

\clearpage

* Correspondence and requests for reprints should be addressed to:

Nancy J. Cox, PhD

900 East 57\textsuperscript{th} Street

KCBD 3300

The University of Chicago

Chicago, IL 60637

Email:
\href{mailto:ncox@medicine.bsd.uchicago.edu}{ncox@medicine.bsd.uchicago.edu}

\bigskip

\bigskip

\bigskip

\bigskip

\bigskip

\bigskip

\bigskip

\bigskip

\bigskip

\bigskip

\bigskip

\bigskip

\bigskip

\clearpage

\textbf{It is widely held that a substantial genetic component underlies
Bipolar Disorder (BD) and other neuropsychiatric disease traits.
\ \ Recent efforts have been aimed at understanding the genetic basis
of disease susceptibility, with genome-wide association studies (GWAS)
unveiling some promising associations. \ Nevertheless, the genetic
etiology of BD remains elusive with a substantial proportion of the
heritability -- which has been estimated to be 80\% based on twin and
family studies -- unaccounted for by the specific genetic variants
identified by large-scale GWAS. \ Furthermore, functional understanding
of associated loci generally lags discovery. \ Studies we report here
provide considerable support to the claim that substantially more
remains to be gained from GWAS on the genetic mechanisms underlying BD
susceptibility, and that a large proportion of the variation in disease
risk may be uncovered through integrative functional genomic
approaches. \ We combine recent analytic advances in heritability
estimation and polygenic modeling and leverage recent technological
advances in the generation of -omics data to evaluate the nature and
scale of the contribution of functional classes of genetic variation to
a relatively intractable disorder. \ We identified
}\textbf{\textit{cis}}\textbf{ eQTLs in cerebellum and parietal cortex
that capture more than half of the total heritability attributable to
SNPs interrogated through GWAS and showed that eQTL-based heritability
estimation is highly tissue-dependent. \ Our findings show that a much
greater resolution may be attained than has been reported thus far on
the number of common loci that capture a substantial proportion of the
heritability to disease risk and that the functional nature of
contributory loci may be clarified }\textbf{\textit{en masse}}\textbf{.
}

\bigskip

Recent advances have facilitated high-throughput explorations of
function-related aspects of the genome
\hyperlink{ENREF1}{\textsuperscript{1-3}}.
\ These developments have yielded comprehensive maps of functional
elements, providing an unparalleled resource to infer the phenotypic
consequences of genetic variation. \ Our group has been particularly
interested in clarifying the impact of regulatory variation on
pathophysiology \hyperlink{ENREF4}{\textsuperscript{4}} and in
integrating heterogeneous genomic datasets to expand on GWAS findings.

For many complex traits, there is an enormous, much-lamented gap
between estimates of heritability derived from population studies
(e.g., family and twin studies) and the proportion of variation
explained by specific genetic loci identified by genome-wide scans
\hyperlink{ENREF5}{\textsuperscript{5}}\textsuperscript{,}\hyperlink{ENREF6}{\textsuperscript{6}}.
\ A second equally fundamental lacuna exists, between the set of
genetic variants identified by GWAS and the functional role of the
implicated variants in mediating the trait. 

Recent reports from our group and others have evaluated the utility of
expression quantitative trait loci (eQTL)
\hyperlink{ENREF4}{\textsuperscript{4}}\textsuperscript{,}\hyperlink{ENREF7}{\textsuperscript{7}}\textsuperscript{,}\hyperlink{ENREF8}{\textsuperscript{8}},
and more recently methylation quantitative trait loci (mQTL)
\hyperlink{ENREF9}{\textsuperscript{9}} and microRNA quantitative trait
loci (miRNAQTL) \hyperlink{ENREF10}{\textsuperscript{10}}, to enhance
discovery of the genetic basis of complex traits. \ All together, these
studies have demonstrated that trait associations from a comprehensive
catalog of published GWAS \hyperlink{ENREF11}{\textsuperscript{11}} as
well as from the top ranks of GWAS results, at least for some complex
traits, are significantly enriched for functional quantitative trait
loci of molecular-level phenotypes, raising the possibility of
utilizing functional genomics to increase the power of a genome-wide
scan to detect true associations as well as \ to clarify the nature of
the identified associations. \ However, it remains to be seen to what
extent the findings emerging from functional genomic studies (e.g.,
eQTLs) may explain the so-called missing heritability characteristic of
GWAS findings. \ \ \ \ 

A flurry of recent studies have utilized random effects models
(linear mixed modeling)
\hyperlink{ENREF12}{\textsuperscript{12}}\textsuperscript{,}\hyperlink{ENREF13}{\textsuperscript{13}}
to estimate the {\textquotedblleft}chip{\textquotedblright}
heritability due to causal loci tagged by SNPs interrogated in GWAS and
a weighted risk score analytic approach (polygenic modeling)
\hyperlink{ENREF14}{\textsuperscript{14}}\textsuperscript{,}\hyperlink{ENREF15}{\textsuperscript{15}}
to quantify the contribution of SNPs that fail to reach genome-wide
significance in GWAS scans. \ However, the findings from the
application of these methods to GWAS thus far have yielded little
insight into the \textit{nature} of the polygenic component to trait
variation. Bipolar Disorder (BD) has been seen as especially
intractable and has contributed little to illuminate the nature of the
genetic architecture of disease risk. \ We hypothesized that the
convergence of functional genomic approaches and these recently
developed techniques for analysis of GWAS data may shed light on the
number of contributory loci for BD and provide a global molecular
perspective on their mode(s) of functional mediation. \ Although a
recent study \hyperlink{ENREF14}{\textsuperscript{14}} concluded that
BD and Schizophrenia share a polygenic component with tens of thousands
of common genetic variants of small effect size likely to be
contributory, studies we report here show that a much greater
resolution may be attained on the number of common loci that capture a
substantial proportion of the heritability to disease risk and that the
functional nature of contributory loci may indeed be clarified
\textit{en masse}. \ \ \ \ 

We had conducted genome-wide expression profiling to map eQTLs (and
mQTLs) in the cerebellum cortex, as previously described
\hyperlink{ENREF9}{\textsuperscript{9}}.
\ \ For the present study, we also utilized the results of eQTL mapping
in parietal cortex. We classified eQTLs into \textit{cis} and
\textit{trans} regulators of gene expression phenotypes on the basis of
the SNP{\textquoteright}s distance to its target gene and on the
strength of the evidence for association with gene expression (see
Methods). \ Using the Wellcome Trust Case Control Consortium (WTCCC)
GWAS study in BD \hyperlink{ENREF16}{\textsuperscript{16}}, we observed
a highly significant enrichment for parietal cortex \textit{cis} eQTLs
and cerebellum \textit{cis} eQTLs (p {\textless} 0.001 for each) among
the top SNPs (defined as p {\textless} 0.001) (Figure 1) relative to
random sets of SNPs (n = 1000) matched on minor allele frequency and
location with respect to nearest gene. \ \ In contrast, no enrichment
was found (p {\textgreater} 0.05) for \textit{cis} eQTLs identified in
lymphoblastoid cell lines (LCLs). \ 

We therefore quantified the contribution of eQTLs in the separate
tissues to the heritability of disease risk. \ Following Yang
\textit{et al}.
\hyperlink{ENREF12}{\textsuperscript{12}},
we utilized a linear mixed model (LMM):

\begin{equation*}
Y = X\beta + g + e
\end{equation*}

\begin{equation*}
var (Y) = A\sigma_{g}^{2} + I\sigma_{e}^{2}
\end{equation*}
where \textbf{Y} is a phenotype vector of size \textit{N }x 1,
\textbf{$\beta $} is a vector of fixed effects, \textbf{g} is a vector
of random additive genetic effects (the {\textquotedblleft}polygenic
component{\textquotedblright} of trait variation) from the set of eQTLs
under study (more generally, any set of QTLs with \textit{a priori}
support from functional genomics, e.g., mQTL
\hyperlink{ENREF9}{\textsuperscript{9}}\textsuperscript{,}\hyperlink{ENREF17}{\textsuperscript{17}}\textsuperscript{,}\hyperlink{ENREF18}{\textsuperscript{18}}
or miRNAQTL\textsuperscript{
}\hyperlink{ENREF10}{\textsuperscript{10}}\textsuperscript{,}\hyperlink{ENREF19}{\textsuperscript{19}}),
and \textbf{e} is a vector of residuals. \ This model leads to two
variance components, namely the additive genetic variance
\textbf{\textit{$\sigma
$}}\textbf{\textit{\textsuperscript{2}}}\textbf{\textit{\textsubscript{g}}}
captured by the tested SNPs and the residual variance
\textbf{\textit{$\sigma
$}}\textbf{\textit{\textsuperscript{2}}}\textbf{\textit{\textsubscript{e}}}.
\ We estimated the (narrow-sense) heritability, \textbf{\textit{$\sigma
$}}\textbf{\textit{\textsuperscript{2}}}\textbf{\textit{\textsubscript{g}}}
/ (\textbf{\textit{$\sigma
$}}\textbf{\textit{\textsuperscript{2}}}\textbf{\textit{\textsubscript{g}}}
+ \textbf{\textit{$\sigma
$}}\textbf{\textit{\textsuperscript{2}}}\textbf{\textit{\textsubscript{e}}}),
explained by the \textit{cis} eQTLs and, separately, the \textit{trans}
eQTLs (despite differential power)
\hyperlink{ENREF20}{\textsuperscript{20}}. \ These estimates, derived
from the genotyped SNPs, quantify the proportion of phenotypic variance
attributable to causal variants in linkage disequilibrium (LD) with the
tested eQTL SNP sets. \ A genetic relationship matrix (GRM), denoted by
\textbf{A} (with entries
\textbf{\textit{A}}\textbf{\textit{\textsubscript{ij}}} between pairs
of individuals \textbf{\textit{i}} and \textbf{\textit{j}}), was
estimated from each SNP set, and variances were estimated using
restricted maximum likelihood (REML)
\hyperlink{ENREF20}{\textsuperscript{20}}. \ \ We were interested in
the heritability captured by the \textit{cis} eQTLs (and the
\textit{trans} eQTLs) as a proportion of the heritability captured by
all SNPs interrogated in the GWAS and passing QC filters (see Methods),
\textbf{\textit{h}}\textbf{\textit{\textsuperscript{2}}}\textbf{\textit{\textsubscript{eqtl
}}}/
\textbf{\textit{h}}\textbf{\textit{\textsuperscript{2}}}\textbf{\textit{\textsubscript{all}}},
hereafter referred to as the \textit{heritability concentration index}.
\ We demonstrated, on theoretical grounds (see Methods) and
empirically, that the heritability concentration index is the same
whether calculated on the observed scale or estimated on the liability
scale (the latter premised on a continuous liability threshold model).
\ Furthermore, it is independent of disease prevalence (\textbf{K}) or
ascertainment bias (\textbf{P}, the proportion of cases in the samples,
which may be {\textgreater}\textbf{K}) (see Methods). \ 

To evaluate the contribution of eQTLs to the (narrow-sense) heritability
of BD, we utilized the TGen+GAIN dataset, which consists of data from
the Bipolar Genome Study (TGen) and from an earlier study (GAIN)
conducted under the GAIN initiative (Table 1). \ All together, the
sample set included in our analysis consists of 2,191 cases and 1,434
controls. \ Selecting samples so that
{\textbar}\textbf{\textit{A}}\textbf{\textit{\textsubscript{ij}}}{\textbar}
{\textless} 0.025 for all pairs \textit{i} and \textit{j }(n = 3,189
individuals) and conducting extensive QC on the genotype data (see
Methods), we performed the heritability estimation analysis in the
TGen+GAIN dataset with and without the study ID as covariate, with no
substantial difference in the estimates. \ Indeed, consistent with a
previous report \hyperlink{ENREF21}{\textsuperscript{21}} on the WTCCC
study of BD \hyperlink{ENREF16}{\textsuperscript{16}}, the genome-wide
SNPs (n {\textgreater} 600,000) explained 35\% (s.e. = 6\%) of the
phenotypic variance on the liability scale (assuming disease prevalence
\textbf{K} = 0.01). \ Remarkably, we found that the cerebellum
\textit{cis} eQTLs (n = 27,107) and parietal cortex \textit{cis} eQTLs
(n = 26,979) included in our analysis had heritability concentration
index
\textbf{\textit{h}}\textbf{\textit{\textsuperscript{2}}}\textbf{\textit{\textsubscript{eqtl
}}}/
\textbf{\textit{h}}\textbf{\textit{\textsuperscript{2}}}\textbf{\textit{\textsubscript{all}}}
of 0.57 and 0.58, respectively. \ The heritability captured by the
\textit{cis} eQTLs previously identified in LCLs
\hyperlink{ENREF4}{\textsuperscript{4}} was less than 1\%, which
demonstrates the tissue dependence of eQTL-based heritability
estimation. \ The inclusion of 20 principal components in the LMM as
fixed effects yielded similar estimates (Supplemental Table 1). \ 

As the results for parietal cortex and cerebellum were found to be
similar (and to significantly differ from the results for LCLs), we
illustrate our approach in the remainder of the paper with the
cerebellum eQTLs unless explicitly stated. \ When we evaluated each
GWAS (TGen and GAIN) separately and partitioned the genetic variance
captured by the \textit{cis} eQTLs onto the individual autosomes (by
fitting the GRMs of the chromosomes simultaneously in a joint
analysis), we found that the estimates of variance explained by each
chromosome (Figure 2) were highly correlated between the two studies
(Pearson correlation = 0.60). \ Furthermore, the signals in the
chromosomal estimates in each GWAS (Figure 2) are not driven by the
large chromosomes. Notably, each GWAS study showed a peak in
heritability estimate on chromosome 11. \ Supplemental Table 2 provides
a list of the chromosome 11 transcripts implicated by this analysis,
including some with a number of independent
(\textbf{\textit{r}}\textbf{\textit{\textsuperscript{2}}} {\textless}
0.10) \textit{cis} eQTLs. \ In particular, using gene set enrichment
analysis \hyperlink{ENREF22}{\textsuperscript{22}}, we found that the
chromosome 11 transcripts with 5 or more \textit{cis} eQTLs
contributing to our estimate of heritability are significantly enriched
for \textit{Immunoglobulin C-2 Type }(IGc2) proteins (Benjamini
Hochberg adjusted p = 6.2 x 10\textsuperscript{{}-3}; Supplemental
Table 3), which raises the question of an infectious/inflammatory etiology to BD
\hyperlink{ENREF23}{\textsuperscript{23}} or a novel CNS mechanism for
this immunoglobulin. \ 

Although our study was not sufficiently powered to reliably identify
\textit{trans} eQTLs, we found that the set of SNPs (n = 6,892)
included in our analysis with association p {\textless} 1.9 x
10\textsuperscript{{}-6} (= 0.05/\# of genes tested) with at least one
gene expression trait in cerebellum had heritability concentration
index of 0.11.

Since experimental or genotyping artifacts as case-control differences
may appear as {\textquotedblleft}heritability{\textquotedblright} (more
so than in the case of quantitative traits, which are less likely to be
correlated with these artifacts)
\hyperlink{ENREF21}{\textsuperscript{21}}, we investigated the effect
of QC thresholds on the estimates of heritability captured by the
genome-wide SNPs as well as the \textit{cis} eQTLs and the
\textit{trans} eQTLs among the interrogated SNPs (see Methods) and
found these estimates to be stable. \ \ 

To test for the presence of any inflation in these estimates, we
conducted heritability estimation using the genome-wide SNPs and,
separately, the subset of \textit{cis} eQTLs on permuted traits
(n=1000) while conditioning on the same corresponding genetic
relationship matrix \textbf{A }(one for each SNP set). \ This analysis
showed that estimates of heritability of simulated phenotypes derived
from the genome-wide SNPs and from the \textit{cis} eQTLs, given their
standard error, were consistent with zero heritability, as we would
expect from a simulated trait with no real association with genotype.
\ Supplemental Figure 1 illustrates the (null) distribution of the
heritability estimate for the \textit{cis} eQTLs. \ \ 

SNP-based heritability estimation is susceptible to confounding by
population stratification \hyperlink{ENREF24}{\textsuperscript{24}}.
The use of principal components (as fixed effects) in the LMM model may
not adequately correct for such confounding, and indeed principal
component-adjusted estimates may yield substantially similar values as
the non-adjusted ones in the presence of fine-scale population
structure \hyperlink{ENREF24}{\textsuperscript{24}}. \ To quantify the
effect of population structure, we estimated heritability from the sum
of heritabilities obtained from two disjoint sets of chromosomes (chr
1-10 and chr 11-22), which we then compared with the estimate derived
from genome-wide SNPs, following Yang \textit{et al}.
\hyperlink{ENREF25}{\textsuperscript{25}}. \ We found that the
heritability concentration index was largely unaffected (0.57 [joint]
vs. 0.55 [disjoint]).

Contributions to the heritability concentration index from causal
variants tagged by the \textit{cis} eQTLs may be distorted by patterns
of LD. \ Indeed, regions of strong LD may amplify a SNP-based estimate
of heritability while contributions from poorly tagged variants may be
underestimated. \ We therefore utilized an LD-adjusted kinship matrix
recently developed by Speed \textit{et al}.
\hyperlink{ENREF26}{\textsuperscript{26}} to tease apart the impact of
local LD from that of the architecture of causal variants on our
estimate of the heritability concentration index. \ We calculated
\textbf{\textit{h}}\textbf{\textit{\textsuperscript{2}}}\textbf{\textit{\textsubscript{eqtl
}}}/
\textbf{\textit{h}}\textbf{\textit{\textsuperscript{2}}}\textbf{\textit{\textsubscript{all}}}
for \textit{cis} eQTLs using a modified genetic relatedness matrix
derived from scaled SNP genotypes. \ The heritability concentration
index did not change substantially with the use of LD-adjusted kinship
matrix (LD-adjusted
\textbf{\textit{h}}\textbf{\textit{\textsuperscript{2}}}\textbf{\textit{\textsubscript{eqtl
}}}/
\textbf{\textit{h}}\textbf{\textit{\textsuperscript{2}}}\textbf{\textit{\textsubscript{all}}}
= 0.572 vs. non-adjusted
\textbf{\textit{h}}\textbf{\textit{\textsuperscript{2}}}\textbf{\textit{\textsubscript{eqtl
}}}/
\textbf{\textit{h}}\textbf{\textit{\textsuperscript{2}}}\textbf{\textit{\textsubscript{all}}}
= 0.570). \ 

The \textit{concentration of heritability} observed for the
\textit{cis} eQTLs was noteworthy. We took a second analytic approach
to evaluate to what extent \textit{cis} eQTLs may have a collective
effect on disease susceptibility. \ The polygenic modeling (PM)
approach has been utilized in several recent studies to demonstrate a
polygenic component to an array of complex human traits, including
schizophrenia \hyperlink{ENREF14}{\textsuperscript{14}}, rheumatoid
arthritis, myocardial infarction and coronary artery disease
\hyperlink{ENREF15}{\textsuperscript{15}}. \ We selected an LD-pruned
set of \textit{cis} eQTLs that meet a P-value threshold in a
{\textquotedblleft}discovery{\textquotedblright} GWAS (TGen) and
calculated a polygenic risk score (see Methods) from this set for each
individual in a {\textquotedblleft}replication{\textquotedblright} GWAS
(GAIN) using the risk alleles and the effects sizes from the
{\textquotedblleft}discovery{\textquotedblright} TGen study. \ We
followed the LD parameters used by Purcell \textit{et al.}
\hyperlink{ENREF14}{\textsuperscript{14}}, to enable direct comparison
(see Methods). \ We utilized several P-value thresholds for association
with BD in TGen (namely, P-value \textsubscript{TGEN} {\textless}
0.0001, 0.01, 0.05, and 0.10) to define a set of \textit{cis} eQTLs and
evaluated the polygenic risk score for association with case-control
status in the second independent GWAS (GAIN) using logistic regression.
\ 

Table 2 illustrates the dependence of the association of the polygenic
risk score with disease status on the P-value \textsubscript{TGEN}
threshold. \ In particular, the set of \textit{cis} eQTLs defined by
P-value \textsubscript{TGEN} {\textless} 0.05 (n = 2,375 SNPs) showed
the most significant association with case-control status. \ Figure 3
shows the chromosomal location of this particular set of \textit{cis}
eQTLs, which appear to be scattered uniformly throughout the genome. \ 

Several observations are worth noting here. \ \ First, this set of eQTLs
constitutes a much smaller number of SNPs (than has previously been
reported) that underlie a polygenic variation in the trait, suggesting
that the use of eQTLs can facilitate unprecedented resolution of the
polygenic component to disease. \ Second, the association of the
polygenic risk score with case-control status is more significant for
the set of \textit{cis} eQTLs with P-value \textsubscript{TGEN}
{\textless} 0.05 than it is for the larger set of interrogated SNPs
that satisfy P-value \textsubscript{TGEN} {\textless} 0.05. \ Third,
our approach not only tests whether a polygenic component predicts
disease risk (as other PM studies do) it also highlights a potential
functional mechanism for the polygenic component. \ Finally, both
statistical approaches, LMM and PM, yield consistent findings on the
effect of \textit{cis} eQTLs, in aggregate, on BD susceptibility.
\ \ \ 

In summary, we have undertaken to quantify the heritability captured
by a functional class of quantitative trait loci for an important
complex disorder. \ Our study not only provides support for the role of
common genetic variation in disease susceptibility, but importantly
also yields a functional basis for the polygenic variation in the
trait. \ This understanding of the genetic architecture of disease risk
could have direct clinical utility and inform the design of future
studies. \ \ 

\clearpage
\textbf{METHODS}

\textit{eQTL Mapping}

The results of our eQTL studies in cerebellum were previously
reported
\hyperlink{ENREF9}{\textsuperscript{9}}.
Here we also present results in parietal cortex. Briefly, DNA and RNA
of cerebellum samples from 153 individuals of European descent were
obtained from the Stanley Medical Research Institute. \ Genotyping was
done on the Affymetrix GeneChip Mapping 5.0 Array (Affymetrix, Santa
Clara, CA, USA). \ Genotype data can be found in the Stanley Genomics
Database. Genome-wide expression profiling was performed using the
Affymetrix Human Gene 1.0 ST Array (\textcolor{black}{GEO
}\textcolor{black}{GSE35974 and GSE35977, for cerebellum and parietal
cortex, respectively)}. \ \ SNPs with call rates less than 99\% were
filtered, as were SNPs that departed from Hardy-Weinberg equilibrium
(P{\textless}0.001) and SNPs with MAF{\textless}10\%. Principal
components analysis, as implemented in Eigenstrat
\hyperlink{ENREF27}{\textsuperscript{27}}, was used to test for the
existence of population structure. \ We performed imputation using MACH
v1.0 \hyperlink{ENREF28}{\textsuperscript{28}}. \ We excluded from
analysis all probes that could be mapped to multiple genome regions as
well as all probes that contain common SNPs (MAF{\textgreater}0.01) on
the basis of 1000 Genomes and HapMap data. \ ComBat
\hyperlink{ENREF29}{\textsuperscript{29}} was used to adjust for batch
effect. Surrogate variable analysis was conducted and identified
surrogate variables were regressed out. Quantile normalization was used
on the residuals. \ Linear regression with dosage of the minor allele
for each SNP was then performed to identify eQTLs. \ In this study, a
\textit{cis} region is defined as within 4 MB of the probe site, while
a \textit{trans} region refers to the rest of the genome. \ A threshold
of {\textless}0.01 was used for the \textit{cis} analysis, while the
\textit{trans} analysis threshold was 0.05 divided by the number of
probes. As our primary interest was in quantifying the heritability
captured by these sets of SNPs (and certain subsets thereof), we
started from these thresholds. \ 

\textit{Genome-wide Association Studies in BD}

Two genome-wide studies of Bipolar Disorder were utilized for our
study of heritability estimation. \ \ The TGen GWAS
\hyperlink{ENREF30}{\textsuperscript{30}} consists of 1,190 cases from
the Bipolar Genome Study and 401 controls. \ In the original GWAS
study, QC procedure excluded SNPs with low MAF ({\textless}1\%),
significant departure from Hardy-Weinberg equilibrium in controls (p
{\textless} 10\textsuperscript{{}-6}), low call rate ({\textless}95\%),
and other criteria. \ A second GWAS, from the GAIN initiative
\hyperlink{ENREF31}{\textsuperscript{31}},
consists of 1,001 cases and 1,033 controls of European descent. \ As in
the TGEN study, SNPs were not included in the analysis if the MAF was
less than 1\%, the SNP violated Hardy-Weinberg equilibrium (p
{\textless} 10\textsuperscript{{}-6}) in control samples, if the call
rate was low ({\textless} 95\%), if there were 3 or more Mendelian
errors, or if there was more than one discrepancy among duplicate
samples. \ 

Following Lee \textit{et al}. \hyperlink{ENREF21}{\textsuperscript{21}},
we performed additional QC steps to ensure the robustness of our
estimates of heritability such as excluding SNPs with p {\textless}
0.05 for Hardy-Weinberg equilibrium and for missingness-difference
between cases and controls. \ Only autosomal SNPs were included in our
heritability estimation analysis.

\textit{Heritability Estimation on the Observed Scale and on the
Liability Scale}

The liability threshold model presupposes an underlying continuous
random variable that defines case-control status. \ Cases are those
subjects for which the liability exceeds a given threshold
\textbf{\textit{t}}. \ For our purposes, suppose that the population
prevalence is \textbf{\textit{K}}. \ Suppose \textbf{\textit{P}} is the
proportion of cases in the sample set; in general, this proportion, an
ascertainment parameter, may not be a random sample from the
population. \ Then the relationship between the heritability on the
observed scale\textbf{\textit{ }}
${h}_{o}^{2}$\textbf{\textit{\textsubscript{ \ }}}and the heritability
on the liability scale  ${h}_{l}^{2}$ is given by the following
expression \hyperlink{ENREF21}{\textsuperscript{21}}:

\begin{equation*}
{h}_{l}^{2}={h}_{0}^{2}\frac{{K}^{2}{(1-K)}^{2}}{P(1-P)}\frac{1}{{\Phi
(t)}^{2}}
\end{equation*}
where \textbf{\textit{$\Phi
$}}\textbf{(}\textbf{\textit{t}}\textbf{)}\textbf{\textit{ }}is the
\textit{y}{}-value of the standard normal curve at the point
\textbf{\textit{t}}. \ \ Note the same scaling factor between the
observed scale and the liability scale applies to the estimate of
heritability for the \textit{cis} eQTLs or \textit{trans} eQTLs as for
the genome-wide SNPs:

\begin{equation*}
\frac{{K}^{2}{(1-K)}^{2}}{P(1-P)}\frac{1}{{\Phi (t)}^{2}}
\end{equation*}
Thus, the ratio of the estimate of heritability attributable to
\textit{cis} eQTLs (or to \textit{trans} eQTLs) relative to the
estimate of heritability attributable to all interrogated SNPs, which
we call eQTL {\textquotedblleft}heritability concentration
index{\textquotedblright}, is the same whether on the observed scale or
on the liability scale, and is independent of disease prevalence and of
ascertainment:

\begin{equation*}
[\frac{{h}_{cis}^{2}}{{h}_{all}^{2}}]_{liability}=[\frac{{h}_{cis}^{2}}{{h}_{all}^{2}}]_{observed}
\end{equation*}
In our heritability estimation analysis, we selected samples so that
{\textbar}\textbf{\textit{A}}\textbf{\textit{\textsubscript{ij}}}{\textbar}
{\textless} 0.025 for all pairs \textit{i} and \textit{j }(leaving n =
3,189 individuals).

\textit{Simulation Analysis Under the Null}

We conducted simulations (n = 1000) to test for the presence of
inflation in our estimates of the heritability captured by the
genome-wide SNPs and, separately, by the \textit{cis} eQTLs. \ In this
analysis, we preserved the genotype correlation structure, utilized the
genetic relationship matrix defined by each set of SNPs, and calculated
the heritability for a permuted trait (n = 1000). \ \ An empirical
p-value was generated for the estimate of heritability, defined as the
number of times the estimate for a simulated trait matches or exceeds
the observed estimate. \ \ Additionally, we determined the number of
times an estimate for a simulated trait is consistent with zero
heritability, e.g., in the case of \textit{cis} eQTLs, the set of
points ( ${h}_{\mathit{cis}}^{2}$ , 
$\mathit{SE}\left({h}_{\mathit{cis}}^{2}\right)$) in [0,1]x[0,1] that
satisfy

\begin{equation*}
{h}_{cis}^2 - 2SE({h}_{cis}^2)\leq 0
\end{equation*}

where  $\mathit{SE}\left({h}_{\mathit{cis}}^{2}\right)$ is the standard
error for the estimate.

\textit{Polygenic Modeling with Functional Variation}

We utilized polygenic modeling
\hyperlink{ENREF14}{\textsuperscript{14}} to evaluate the effect of
large numbers of weakly associated SNPs characterized by very small
allele frequency differences between cases and controls. To facilitate
direct comparison with the Purcell \textit{et al.} study, we pruned a
given SNP set (e.g., the \textit{cis} eQTLs) to filter SNPs in strong
LD with other SNPs (using a pairwise
\textit{r}\textit{\textsuperscript{2}} of 0.25, within a 200-SNP
sliding window). \ \ Using the set of risk alleles from the resulting
LD-pruned set and the corresponding effect size from a
{\textquotedblleft}discovery{\textquotedblright} GWAS, we calculated,
in a {\textquotedblleft}validation{\textquotedblright} GWAS, a
polygenic score from the log odds ratio-weighted sum of risk allele
count  ${x}_{i,j}$\textbf{ }for each individual \textbf{\textit{j}}: \ 

\begin{equation*}
{S}_{j}=\sum _{i}{\log ({OR}_{i,j}){x}_{i,j}}
\end{equation*}

For polygenic modeling, we evaluated the sets of \textit{cis} eQTLs
defined by P-value \textsubscript{TGEN} {\textless} 0.0001, 0.01, 0.05,
and 0.10. \ We tested the calculated polygenic score for association
with disease status.

\bigskip

\clearpage
\textbf{Figure 1}. \textbf{Enrichment of eQTLs among the top SNPs from
the WTCCC study. \ }The highest ranked SNPs (p{\textless}0.001) in the
WTCCC study of BD were found to be enriched for cerebellum \textit{cis}
eQTLs and parietal cortex \textit{cis} eQTLs relative to random SNPs
matched on minor allele frequency and location with respect to nearest
gene. \ The black dot represents the observed count and the histogram
depicts the empirical (null) distribution of the eQTL count generated
from the randomly drawn SNPs. \ \ (In contrast, no evidence for eQTL
enrichment was observed in LCLs.)

\bigskip

\includegraphics[width=5in,height=5in]{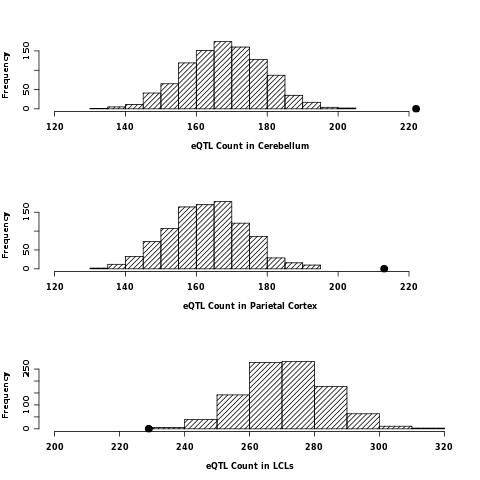}
\textbf{ }

\clearpage
\textbf{Figure 2. Partitioning of variance captured by
}\textbf{\textit{cis}}\textbf{ eQTLs by chromosome. \ }The estimates of
variance, from the two GWAS GAIN and TGen, captured by the cerebellum
\textit{cis} eQTLs on each chromosome were highly correlated (Pearson
correlation = 0.60). \ Red corresponds to the GAIN study while orange
the TGen study. \ The estimates shown here are on the observed scale.

\bigskip

\includegraphics[width=5in,height=5in]{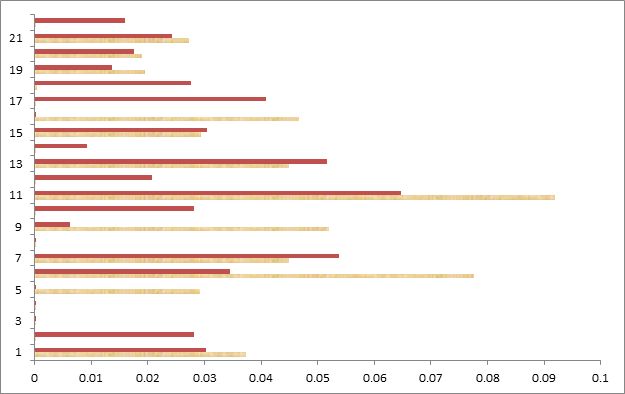}

\bigskip

\clearpage
\textbf{Figure 3. \ eQTL-based polygenic modeling. \ }We selected an
LD-pruned set of \textit{cis} eQTLs that meet a P-value threshold in a
{\textquotedblleft}discovery{\textquotedblright} GWAS (TGen) -- P-value
\textsubscript{TGEN} {\textless} 0.0001, 0.01, 0.05, and 0.10 were
tested -- and calculated a polygenic risk score from this set for each
individual in a {\textquotedblleft}replication{\textquotedblright} GWAS
(GAIN) using the risk alleles and the effects sizes from the
{\textquotedblleft}discovery{\textquotedblright} TGen study. \ The set
of \textit{cis} eQTLs defined by P-value \textsubscript{TGEN}
{\textless} 0.05 (n = 2,375 SNPs) showed the most significant
association with case-control status in GAIN using logistic regression.
Each red mark indicates a \textit{cis} eQTL SNP included in the
polygenic model.\textbf{ }

\includegraphics[width=6.2638in,height=6.4874in]{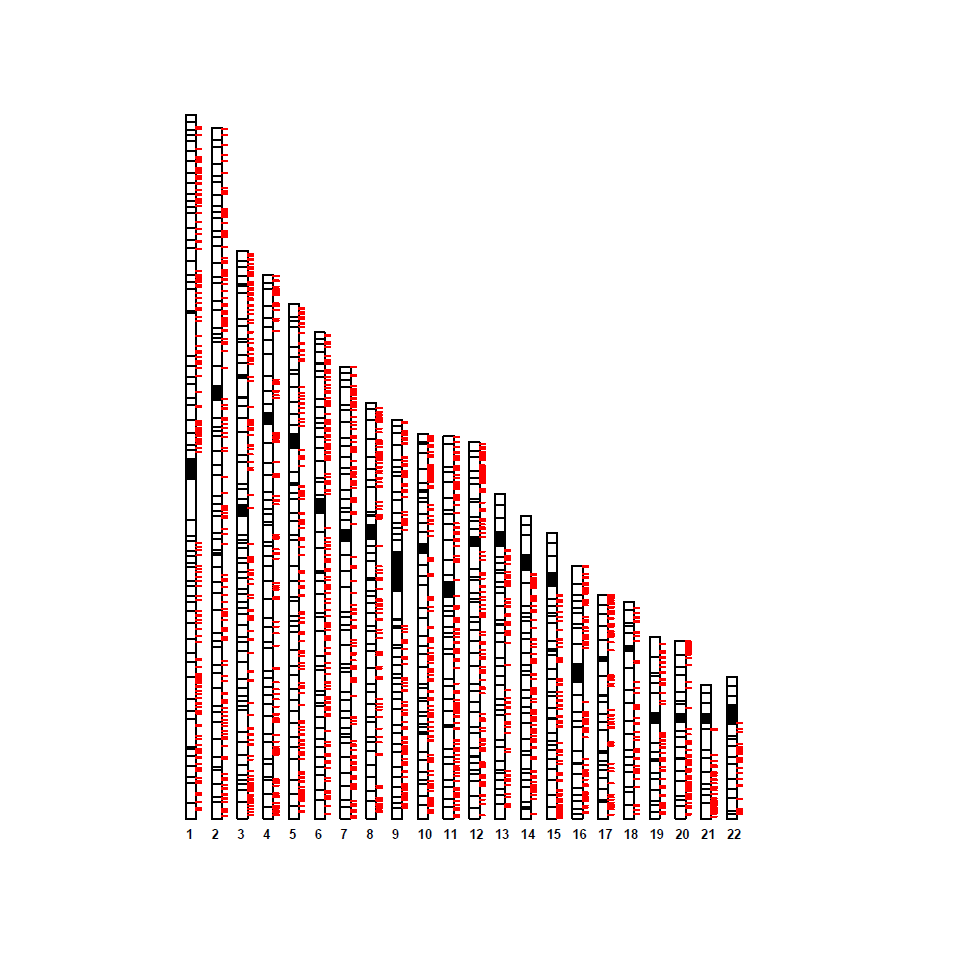}

\bigskip

\clearpage
\textbf{Table 1. \ Genome-wide association studies of Bipolar Disorder
evaluated in our study.}

\begin{flushleft}
\tablehead{}
\begin{supertabular}{|m{2.05316in}|m{2.05316in}|m{2.05316in}|}
\hline
\textbf{Study} &
\textbf{Cases} &
\textbf{Controls}\\\hline
TGEN &
1190 &
401\\\hline
GAIN &
1001 &
1033\\\hline
\end{supertabular}
\end{flushleft}

\bigskip

\bigskip

\clearpage
\textbf{Table 2. \ Polygenic modeling with eQTLs. }Using the
\textit{cis} eQTLs, we determined, for each individual in the
{\textquotedblleft}replication{\textquotedblright} GAIN study, a
polygenic risk score, which is defined as the sum of the risk allele
counts weighted by the log odds ratio using the risk alleles and the
effect sizes from the {\textquotedblleft}discovery{\textquotedblright}
TGen study. \ The polygenic risk score was then tested for association
with disease status in the GAIN study. \ An LD-pruned SNP set
consisting of \textit{cis} eQTLs with p {\textless} 0.05 in the TGEN
study showed the most significant association with case control status
in the GAIN study. \ 

\begin{flushleft}
\tablehead{}
\begin{supertabular}{|m{2.37126in}|m{4.14836in}|}
\hline
\textbf{TGen p-value threshold} &
\textbf{p-value of association of polygenic score with disease
status}\\\hline
\begin{flushleft}
\begin{tabular}{m{2.1712599in}}
\textcolor{black}{p {\textless} 0.0001}\\
\textcolor{black}{p {\textless} 0.01}\\
\textcolor{black}{p {\textless} 0.05}\\
\textcolor{black}{p {\textless} 0.1}\\
\end{tabular}
\end{flushleft}
~
 &
\begin{flushleft}
\begin{tabular}{m{3.94906in}}
\raggedleft\arraybslash \textcolor{black}{0.894}\\
\raggedleft\arraybslash \textcolor{black}{0.0245}\\
\raggedleft\arraybslash \textcolor{black}{0.01}\\
\raggedleft\arraybslash \textcolor{black}{0.0115}\\
\end{tabular}
\end{flushleft}
~
\\\hline
\end{supertabular}
\end{flushleft}

\bigskip

\clearpage
\textbf{ACKNOWLEDGMENTS}

This work was funded by the Genotype-Tissue Expression project (GTeX)
(R01 MH090937) and PAAR (Pharmacogenetics of Anti-cancer Agents
Research; U01 GM61393).

\bigskip

\bigskip

\textbf{AUTHOR CONTRIBUTIONS}

ERG and NJC conceived and designed the study. \ ERG wrote the
manuscript. \ NJC and DLN supervised the study. \ ERG, HKI, CL, and DLN
contributed reagents/materials/analysis tools. \ The BiGS consortium
participated in patient diagnosis and sample collection. \ All authors
edited and approved the manuscript.

\clearpage
\textbf{REFERENCES}

1.\ \ Wang, Z., Gerstein, M. \& Snyder, M. RNA-Seq: a revolutionary tool
for transcriptomics. \textit{Nat Rev Genet} \textbf{10}, 57-63 (2009).

2.\ \ Degner, J.F. et al. DNase I sensitivity QTLs are a major
determinant of human expression variation. \textit{Nature}
\textbf{482}, 390-4 (2012).

3.\ \ Ward, L.D. \& Kellis, M. HaploReg: a resource for exploring
chromatin states, conservation, and regulatory motif alterations within
sets of genetically linked variants. \textit{Nucleic Acids Res}
\textbf{40}, D930-4 (2012).

\hypertarget{ENREF4}{}4.\ \ Nicolae, D.L. et al. Trait-associated SNPs
are more likely to be eQTLs: annotation to enhance discovery from GWAS.
\textit{PLoS Genet} \textbf{6}, e1000888 (2010).

\hypertarget{ENREF5}{}5.\ \ Manolio, T.A. et al. Finding the missing
heritability of complex diseases. \textit{Nature} \textbf{461}, 747-53
(2009).

\hypertarget{ENREF6}{}6.\ \ Eichler, E.E. et al. Missing heritability
and strategies for finding the underlying causes of complex disease.
\textit{Nat Rev Genet} \textbf{11}, 446-50 (2010).

\hypertarget{ENREF7}{}7.\ \ Gamazon, E.R., Huang, R.S., Cox, N.J. \&
Dolan, M.E. Chemotherapeutic drug susceptibility associated SNPs are
enriched in expression quantitative trait loci. \textit{Proc Natl Acad
Sci U S A} \textbf{107}, 9287-92 (2010).

\hypertarget{ENREF8}{}8.\ \ Nica, A.C. et al. Candidate causal
regulatory effects by integration of expression QTLs with complex trait
genetic associations. \textit{PLoS Genet} \textbf{6}, e1000895 (2010).

\hypertarget{ENREF9}{}9.\ \ Gamazon, E.R. et al. Enrichment of
cis-regulatory gene expression SNPs and methylation quantitative trait
loci among bipolar disorder susceptibility variants. \textit{Mol
Psychiatry} (2012).

\hypertarget{ENREF10}{}10.\ \ Gamazon, E.R., Ziliak, D., Im, H.K.,
\ LaCroix B., Park, D.S., Cox N.J., Huang, R.S. Genetic Architecture of
microRNA Expression: Implications for the Transcriptome and Complex
Traits. \textit{American Journal of Human Genetics} (2012).

\hypertarget{ENREF11}{}11.\ \ Hindorff, L.A. et al. Potential etiologic
and functional implications of genome-wide association loci for human
diseases and traits. \textit{Proc Natl Acad Sci U S A} \textbf{106},
9362-7 (2009).

\hypertarget{ENREF12}{}12.\ \ Yang, J. et al. Common SNPs explain a
large proportion of the heritability for human height. \textit{Nat
Genet} \textbf{42}, 565-9 (2010).

\hypertarget{ENREF13}{}13.\ \ Visscher, P.M., Yang, J. \& Goddard, M.E.
A commentary on {\textquotesingle}common SNPs explain a large
proportion of the heritability for human height{\textquotesingle} by
Yang et al. (2010). \textit{Twin Res Hum Genet} \textbf{13}, 517-24
(2010).

\hypertarget{ENREF14}{}14.\ \ Purcell, S.M. et al. Common polygenic
variation contributes to risk of schizophrenia and bipolar disorder.
\textit{Nature} \textbf{460}, 748-52 (2009).

\hypertarget{ENREF15}{}15.\ \ Stahl, E.A. et al. Bayesian inference
analyses of the polygenic architecture of rheumatoid arthritis.
\textit{Nat Genet} \textbf{44}, 483-9 (2012).

\hypertarget{ENREF16}{}16.\ \ Genome-wide association study of 14,000
cases of seven common diseases and 3,000 shared controls.
\textit{Nature} \textbf{447}, 661-78 (2007).

\hypertarget{ENREF17}{}17.\ \ Gibbs, J.R. et al. Abundant quantitative
trait loci exist for DNA methylation and gene expression in human
brain. \textit{PLoS Genet} \textbf{6}, e1000952 (2010).

\hypertarget{ENREF18}{}18.\ \ Bell, J.T. et al. DNA methylation patterns
associate with genetic and gene expression variation in HapMap cell
lines. \textit{Genome Biol} \textbf{12}, R10 (2011).

\hypertarget{ENREF19}{}19.\ \ Ziebarth, J.D., Bhattacharya, A., Chen, A.
\& Cui, Y. PolymiRTS Database 2.0: linking polymorphisms in microRNA
target sites with human diseases and complex traits. \textit{Nucleic
Acids Res} \textbf{40}, D216-21 (2012).

\hypertarget{ENREF20}{}20.\ \ Yang, J., Lee, S.H., Goddard, M.E. \&
Visscher, P.M. GCTA: a tool for genome-wide complex trait analysis.
\textit{Am J Hum Genet} \textbf{88}, 76-82 (2011).

\hypertarget{ENREF21}{}21.\ \ Lee, S.H., Wray, N.R., Goddard, M.E. \&
Visscher, P.M. Estimating missing heritability for disease from
genome-wide association studies. \textit{Am J Hum Genet} \textbf{88},
294-305 (2011).

\hypertarget{ENREF22}{}22.\ \ Huang da, W., Sherman, B.T. \& Lempicki,
R.A. Systematic and integrative analysis of large gene lists using
DAVID bioinformatics resources. \textit{Nat Protoc} \textbf{4}, 44-57
(2009).

\hypertarget{ENREF23}{}23.\ \ Hinze-Selch, D. Infection, treatment and
immune response in patients with bipolar disorder versus patients with
major depression, schizophrenia or healthy controls. \textit{Bipolar
Disord} \textbf{4 Suppl 1}, 81-3 (2002).

\hypertarget{ENREF24}{}24.\ \ Browning, S.R. \& Browning, B.L.
Population structure can inflate SNP-based heritability estimates.
\textit{Am J Hum Genet} \textbf{89}, 191-3; author reply 193-5 (2011).

\hypertarget{ENREF25}{}25.\ \ Yang, J. et al. Genome partitioning of
genetic variation for complex traits using common SNPs. \textit{Nat
Genet} \textbf{43}, 519-25 (2011).

\hypertarget{ENREF26}{}26.\ \ Speed, D., Hemani, G., Johnson, M.R. \&
Balding, D.J. Improved Heritability Estimation from Genome-wide SNPs.
\textit{Am J Hum Genet} \textbf{91}, 1011-21 (2012).

\hypertarget{ENREF27}{}27.\ \ Price, A.L. et al. Principal components
analysis corrects for stratification in genome-wide association
studies. \textit{Nat Genet} \textbf{38}, 904-9 (2006).

\hypertarget{ENREF28}{}28.\ \ Li, Y., Willer, C.J., Ding, J., Scheet, P.
\& Abecasis, G.R. MaCH: using sequence and genotype data to estimate
haplotypes and unobserved genotypes. \textit{Genet Epidemiol}
\textbf{34}, 816-34 (2010).

\hypertarget{ENREF29}{}29.\ \ Johnson, W.E., Li, C. \& Rabinovic, A.
Adjusting batch effects in microarray expression data using empirical
Bayes methods. \textit{Biostatistics} \textbf{8}, 118-27 (2007).

\hypertarget{ENREF30}{}30.\ \ Smith, E.N. et al. Genome-wide association
of bipolar disorder suggests an enrichment of replicable associations
in regions near genes. \textit{PLoS Genet} \textbf{7}, e1002134 (2011).

31.\ \ Smith, E.N. et al. Genome-wide association study of bipolar
disorder in European American and African American individuals.
\textit{Mol Psychiatry} \textbf{14}, 755-63 (2009).

\bigskip

\bigskip

\bigskip

\clearpage
\textbf{Members of the Bipolar Disorder Genome Study (BiGS) Consortium}

John R Kelsoe\textsuperscript{1,2}, Tiffany A
Greenwood\textsuperscript{1}, Caroline M Nievergelt\textsuperscript{1},
Thomas B Barrett\textsuperscript{1}, Rebecca
McKinney\textsuperscript{1}, Paul D Shilling\textsuperscript{1},
Nicholas J Schork\textsuperscript{3--5}, Erin N
Smith\textsuperscript{3,4}, Cinnamon S Bloss\textsuperscript{3,5}, John
Nurnberger\textsuperscript{6}, Howard J Edenberg\textsuperscript{7,8},
Tatiana Foroud\textsuperscript{8}, Daniel L Koller\textsuperscript{6},
William Scheftner\textsuperscript{9}, William B
Lawson\textsuperscript{10}, Evaristus A Nwulia\textsuperscript{10},
Maria Hipolito\textsuperscript{10}, William
Coryell\textsuperscript{11}, John Rice\textsuperscript{12}, William
Byerley\textsuperscript{13}, Francis McMahon\textsuperscript{14}, David
TW Chen\textsuperscript{14}, Thomas G Schulze\textsuperscript{14,15},
Wade Berrettini\textsuperscript{16}, James B
Potash\textsuperscript{17,18}, Peter P Zandi\textsuperscript{17},
Pamela B Mahon\textsuperscript{17}, Melvin McInnis\textsuperscript{19},
David Craig\textsuperscript{20} and Szabolcs
Szelinger\textsuperscript{20}

\textsuperscript{1}Department of Psychiatry, University of California,
San Diego, CA, USA; \textsuperscript{2}Department of Psychiatry, VA San
Diego Healthcare System, La Jolla, CA, USA; \textsuperscript{3}Scripps
Genomic Medicine, Scripps Translational Science Institute, La Jolla,
CA, USA; \textsuperscript{4}Department of Molecular and Experimental
Medicine, The Scripps Research Institute, La Jolla, CA, USA;
\textsuperscript{5}Scripps Health, La Jolla, CA, USA;
\textsuperscript{6}Department of Psychiatry, Indiana University School
of Medicine, Indianapolis, IN, USA; \textsuperscript{7}Department of
Biochemistry and Molecular Biology, Indiana University School of
Medicine, Indianapolis, IN, USA; \textsuperscript{8}Department of
Medical and Molecular Genetics, Indiana University School of Medicine,
Indianapolis, IN, USA; \textsuperscript{9}Department of Psychiatry,
Rush University, Chicago, IL, USA; \textsuperscript{10}Department of
Psychiatry, Howard University, Washington, DC, USA;
\textsuperscript{11}Department of Psychiatry, University of Iowa, Iowa
City, IA, USA; \textsuperscript{12}Division of Biostatistics,
Washington University, St Louis, MO, USA;
\textsuperscript{13}Department of Psychiatry, University of California,
San Francisco, San Francisco, CA, USA; \textsuperscript{14}Human
Genetics Branch, National Institute of Mental Health Intramural
Research Program, National Institutes of Health, US Department of
Health and Human Services, Bethesda, MD, USA;
\textsuperscript{15}Section on Psychiatric Genetics, Department of
Psychiatry and Psychotherapy, Georg-August-University, G\"ottingen,
Germany; \textsuperscript{16}Department of Psychiatry, University of
Pennsylvania, Philadelphia, PA, USA; \textsuperscript{17}Department of
Psychiatry, Johns Hopkins School of Medicine, Baltimore, MD, USA;
\textsuperscript{18}Department of Psychiatry, University of Iowa, Iowa
City, IA, USA; \textsuperscript{19}Department of Psychiatry, University
of Michigan, Ann Arbor, MI, USA and \textsuperscript{20}Neurogenomics
Division, The Translational Genomics Research Institute, Phoenix, AZ,
USA.

\bigskip

\bigskip
\end{document}